\documentclass[11pt]{article}
\usepackage{amsmath,amssymb,color,graphics,epsfig,cite}

\usepackage{amsfonts}

\newcommand{\be}{\begin{equation}}
\newcommand{\ee}{\end{equation}}
\newcommand{\bea}{\setlength\arraycolsep{2pt} \begin{eqnarray}}
\newcommand{\eea}{\end{eqnarray}}
\newcommand{\nn}{\nonumber}

\def\ft#1#2{{\textstyle{\frac{\scriptstyle #1}{\scriptstyle #2} } }}
\def\fft#1#2{{\frac{#1}{#2}}}

\def\0{{\sst{(0)}}}
\def\1{{\sst{(1)}}}
\def\2{{\sst{(2)}}}
\def\3{{\sst{(3)}}}
\def\4{{\sst{(4)}}}
\def\5{{\sst{(5)}}}
\def\6{{\sst{(6)}}}
\def\7{{\sst{(7)}}}
\def\8{{\sst{(8)}}}
\def\sst#1{{\scriptscriptstyle #1}}

\begin{document}

\begin{center}
{\Large {\bf A Correspondence between Ricci-flat Kerr\\
and Kaluza-Klein AdS Black Hole}}

\vspace{20pt}

{\large Liang Ma and H. L\"u}

\vspace{10pt}

{\it Center for Joint Quantum Studies and Department of Physics,\\
School of Science, Tianjin University, Tianjin 300350, China }

\vspace{40pt}

\underline{ABSTRACT}
\end{center}

We establish an explicit correspondence of Einstein gravity on the squashed spheres that are the $U(1)$ bundles over $\mathbb{CP}^m$ to the Kaluza-Klein AdS gravity on the tori.  This allows us to map the Ricci-flat Kerr metrics in odd dimensions with all equal angular momenta to charged Kaluza-Klein AdS black holes that can be lifted to become singly rotating M-branes and D3-branes. Furthermore, we find maps between Ricci-flat gravitational instantons to the AdS domain walls. In particular the supersymmetric bolt instantons correspond to domain walls that can be interpreted as distributed M-branes and D3-branes, whilst the non-supersymmetric Taub-NUT solutions yield new domain walls that can be lifted to become solutions in M-theory or type IIB supergravity. The correspondence also inspires us to obtain a new superpotential in the Kaluza-Klein AdS gravity in four dimensions.

\vfill{\footnotesize  liangma@tju.edu.cn \ \ \ mrhonglu@gmail.com}


\thispagestyle{empty}
\pagebreak

\tableofcontents
\addtocontents{toc}{\protect\setcounter{tocdepth}{2}}


\section{Introduction}

Duality symmetries or correspondences provide powerful tools to study a theory in the regions that may be otherwise unaccessible. The most noted example is the electromagnetic duality conjecture that states the whole physics, including both the perturbative and non-perturbative regions, is the same under the Hodge dual of the Maxwell field, together with turning the coupling to its inverse. This duality symmetry is realized within the same spacetime dimensions. String theory provides a plethora of new dualities. These typically manifest themselves in the lower-energy effective theories as global symmetries or duality field transformations in the Kaluza-Klein reduced theories. For example, the type IIA/IIB supergravities become the same after reduced on a circle, {\it via} appropriate field redefinitions \cite{Bergshoeff:1995as,Cvetic:1999zs}. The Cremmer-Julia groups of maximal supergravities \cite{Cremmer:1978ds,Julia:1980gr,Cremmer:1997ct} are promoted to U-dualities in string and M-theory \cite{Hull:1994ys}.  The celebrated AdS/CFT correspondence \cite{Maldacena:1997re} is the different type, relating weakly-coupled gravity in the anti-de Sitter (AdS) background to some strongly coupled conformal field theory (CFT) on its boundary, an example of the alternatives to compactification \cite{Randall:1999vf}

In this paper, we study a new type of correspondence proposed in \cite{Caldarelli:2013aaa}.  (See also \cite{Caldarelli:2012hy,Caldarelli:2018azk}.) It relates pure Einstein gravity on a sphere to Einstein gravity with a negative cosmological constant $\Lambda$ on a suitable torus. The correspondence can be easily understood from the vacuum structure. The Minkowski spacetime is conformal to AdS warped with a sphere.  Specifically, in $\hat D= p+2 + n$ dimensions, the Minkowski vacuum can be expressed as
\be
d\hat s_{\hat D}^2=\fft{r^2}{\ell^2} \Big(ds_{p+2}^2 + \ell^2 d\Omega_n^2\Big)\,,\qquad
ds_{p+2}^2 = \fft{\ell^2}{r^2} \Big(-dt^2 + dx^i dx^i + dr^2 \Big),
\ee
where $i=1,2,\ldots, p$.  This can be mapped to the AdS vacuum of Einstein gravity in $\widetilde D=p+2+q$ dimensions on torus ${\cal T}^q$, with
\be
d\tilde s_{\widetilde D}^2 = ds_{p+2}^2 + \fft{\ell^2}{r^2} ds_{{\cal T}^q}^2\,.
\ee
At the level of these vacua, there is no obvious constrain on the dimensional parameters.  Performing the Kaluza-Klein reductions, keeping only the breathing mode of each theory, it was shown \cite{Caldarelli:2013aaa} that the two theories in the lower $(p+2)$-dimensions can map from one to the other, provided that the dimensional parameters must satisfy $n\leftrightarrow -(\widetilde D-2)$ or equivalently $q\leftrightarrow - (\hat D-2)$. Since the duality maps the $r\rightarrow 0$ region of Minkowski to the asymptotic boundary of the AdS, {\it via} $r\rightarrow \ell^2/r$, it raises immediate objection that the $r\rightarrow 0$ is the spacetime singularity of the Ricci-flat Schwarzschild black hole. However, the negative dimension involved resolves the puzzle in that in negative spacetime dimensions, $r\rightarrow 0$ gives formerly the ``Minkowski spacetime'', while $r\rightarrow \infty$ gives the singularity. On the other hand, the negative dimensions involved in the map indicates that there is no physical internal dimensions involved, even though the Kaluza-Klein idea is essential in the discussion of this AdS/Ricci-flat correspondence.  Thus this correspondence is a different alternative to compactification.

In this paper, we would like to extend the discussion of \cite{Caldarelli:2013aaa} to include matter fields, such as the Maxwell field. However, the naive idea of simply adding a Maxwell field in both theories will break the correspondence.  In string theory, analogous duality transformations typically interchange the matter fields with geometric fields that arise from the internal spaces.  For example, the $T$-duality of string theory interchanges the string winding modes, which are matter, with the Kaluza-Klein modes, which are geometrical.  In fact the AdS/Ricci-flat correspondence discussed above indeed interchanges the Ricci scalar $R_n$ of the $n$-sphere with the negative cosmological constant $\Lambda$, which can be viewed as matter.

We thus expect that the Maxwell field should arise from the metric as a Kaluza-Klein vector in one theory.  We note that in odd $n=2m+1$ dimensions, the sphere can be expressed as a $U(1)$ bundle over $\mathbb{CP}^m$ and the Maxwell field can naturally reside in the fibre without breaking the isometry group.  We thus consider pure Einstein gravity on squashed sphere that is the $U(1)$ bundle over $\mathbb{CP}^m$.  We find this theory corresponds to a special class of Einstein-Maxwell-Dilaton (EMD) theory with a nontrivial scalar potential whose extremum gives the AdS vacuum.  We identify that the theory in lower dimensions can be embedded in gauged supergravities and can be called the Kaluza-Klein AdS gravity.  This allows us to make a map between Ricci-flat Kerr metrics with all equal angular momenta in odd dimensions to charged Kaluza-Klein AdS black holes.

The paper is organized as follows.  In section 2, we present the two theories and establish the correspondence by finding the explicit duality transformations. In section 3, we present the correspondence between the Ricci-flat Kerr metrics and charged AdS black holes.  We further establish the map between Ricci-flat gravitational instantons to domain walls, for both the supersymmetric and non-supersymmetric cases.  We conclude the paper in section 4.  In appendix A, we present the details of the Kaluza-Klein reduction on the $U(1)$ bundles over $\mathbb{CP}^m$.  In appendix B, we present the equations of domain wall solutions.  In appendix C, we present equations of the gravitational instantons whose level surfaces are the $U(1)$ bundles over $\mathbb{CP}^m$.

\section{An AdS/Ricci-flat correspondence}
\label{sec:mapoftheories}

\subsection{The two theories}

In this section, we establish the explicit transformation rules between two theories.  One is simply the Einstein theory of pure gravity in general $\hat D$ dimensions, and the action is
\be
\hat S = \fft{1}{16\pi \hat G} \int d^{\hat D} x \hat {\cal L}\,,\qquad \hat {\cal L}=\sqrt{|\hat g|}
\hat R\,,\label{puregrav}
\ee
where $\hat G$ is the Newton's constant.

The other is a specific EMD-AdS theory in general $\widetilde D$ dimensions, involving the metric, a Maxwell field $A_\mu$ and a dilaton-like scalar.  The action is
\bea
\widetilde S &=& \fft1{16\pi \widetilde G} \int d^{\widetilde D} x \widetilde {\cal L}\,,\nn\\
\widetilde {\cal L} &=& \sqrt{|\tilde g|}  \Big(\widetilde R - \ft12(\partial\phi)^2 -V(\phi)- \fft{1}{4}e^{\sqrt{\fft{2(\widetilde D-1)}{\widetilde D-2}}\phi} F^2\Big),\label{kkemdads}
\eea
where $F=dA$ and the scalar potential can be expressed in terms of a superpotential
\bea
V &=& \Big(\fft{dW}{d\phi}\Big)^2 - \fft{\tilde D-1}{2(\tilde D-2)} W^2\,,\nn\\
W &=& \fft1{\ell \sqrt2} \Big((\tilde D-3)
e^{-\fft{(\tilde D-1)\phi}{\sqrt{(\tilde D-1)(\tilde D-2)}}} + (\tilde D-1) e^{\fft{(\tilde D-3)\phi}{
\sqrt{2(\tilde D-1)(\tilde D-2)}}}\Big).\label{superpot}
\eea
The scalar potential has a stationary point $\phi=0$, giving a cosmological constant, with
\be
\Lambda=\ft12 V(0) = -\fft{(\widetilde D-1)(\widetilde D-2)}{2\ell^2}\,,\qquad W(0)=\fft{\sqrt2 (\widetilde D-2)}{\ell}\,.
\ee
This EMD-AdS theory is a special class of Lagrangians considered in \cite{Gao:2004tu,Lu:2013eoa}, where the superpotential was obtained in \cite{Lu:2013eoa}. When $\widetilde D=4,5,6,7$, the theory can be obtained from consistent truncations of the corresponding maximal gauged supergravities. (See, e.g.~\cite{Cvetic:1999xp}.) Turning off the cosmological constant, we obtain the Kaluza-Klein theory from $S^1$ reduction of pure gravity and hence it can be also referred to as the Kaluza-Klein AdS gravity.  Notable solutions include the dyonic AdS black hole in $\widetilde D=4$ \cite{Lu:2013ura} and charged rotating black holes in general $\widetilde D$ \cite{Wu:2011zzh}.

For later purposes, we find it is convenient to introduce $\Phi$, given by
\be
\Phi=\exp \Big(\fft{\phi}{\sqrt{2(\widetilde D-1)(\widetilde D-2)}}\Big)\,.
\ee
The Lagrangian (\ref{kkemdads}) becomes
\be
\widetilde {\cal L} = \sqrt{|\tilde g|}  \Big(\widetilde R - \tilde n(\tilde n-1) \Phi^{-2} (\partial \Phi)^2 -\fft{1}{\ell^2 \Phi^2}\Big((\tilde n-1){\Phi^{-2\tilde n}} -(\tilde n^2-1)\Big)- \fft{1}{4}\Phi^{-2(\tilde n-1)} F^2\Big),\label{kkemdads2}
\ee
where $\tilde n=-(\widetilde D-2)$.

\subsection{The correspondence}

As we remarked in the previous subsection, if we turn off the cosmological constant ($\ell\rightarrow \infty$), the Lagrangian (\ref{kkemdads}) can be obtained from pure gravity (\ref{puregrav}) {\it via} the Kaluza-Klein reduction on $S^1$. Such connection is not available for non-vanishing scalar potential.  In the case when the theory can be embedded in gauged supergravities, the theory can be obtained from the Kaluza-Klein sphere reductions of $D=11$ or $D=10$ supergravities, where the electric charge carried by the Maxwell field has an origin of angular momentum of the singly rotating M2/5-brane or D3-brane \cite{Cvetic:1999xp}.

In this section, we shall establish a correspondence map between the two seemingly different theories (\ref{puregrav}) and (\ref{kkemdads}). We follow the general idea of \cite{Caldarelli:2013aaa}, namely one can establish a map between Einstein gravity reduced on sphere and the Kaluza-Klein AdS gravity reduced on torus, where the Maxwell field from the EMD theory corresponds to some specific Kaluza-Klein vector of the pure gravity.  In order to do this, we consider general odd-dimensional spheres which can be expressed as $U(1)$ bundles over some complex projective $\mathbb{CP}$ space.  In other words, the metric for the unit round sphere in odd dimensions $n=2m+1$ can be written as
\be
d\Omega^2_{n} = \sigma^2 + d\Sigma_{m}^2\,,\qquad \sigma=d\psi + B^{(m)}\,,\qquad n=2m+1\,,
\ee
where $J=2dB^{(m)}$ is the K\"ahler 2-form of the $\mathbb{CP}^m$, whose metric is $d\Sigma_{m}^2$, of $2m$ real dimensions.  In this paper, we follow exactly the same conventions of \cite{Cvetic:2018ipt} for the $\mathbb{CP}^m$ metric and its K\"ahler potential and therefore we shall not repeat the details here. The $\hat D=p+2 + n$ dimensional reduction ansatz is
\be
d\hat s_{\hat D}^2=X^2 \Bigg(ds_{p+2}^2 + \ell^2 \Big(U^{-2(n-1)}(\sigma + \ell^{-1} {\cal A})^2 + U^2 d\Sigma_{m}^2\Big)\Bigg),\qquad {\cal A}={\cal A}_\mu dx^\mu\,.\label{reduction1}
\ee
Here the scalar quantities $(X,U)$ and vector ${\cal A}_\mu$ are all functions of the coordinates associated with the lower-dimensional spacetime metric $ds_{p+2}^2$.  The function $X$ describes the breathing mode of the $S^n$ and $U$ parameterizes the squashing of the $S^n$, while keeping its volume fixed.  Since the reduction ansatz contains all the singlets of the isometry of the $U(1)$ bundle of $\mathbb{CP}^m$, it follows that the reduction must be consistent. The lower-dimensional Lagrangian is
\bea
{\cal L}_{p+2} &=& \ell^n \sqrt{|g|} X^{\hat D-2} \Bigg(R + (\hat D-1)(\hat D-2) X^{-2} (\partial X)^2-n(n-1) U^{-2} (\partial U)^2\nn\\
&&-\fft{1}{\ell^2 U^2}\Big((n-1) U^{-2n} -(n^2-1)\Big)-\fft14 U^{-2(n-1)}{\cal F}^2\Bigg),\label{lowd1}
\eea
where ${\cal F}=d{\cal A}$. The Newton's constant in $(p+2)$ dimensions becomes
\be
G_{p+2} =  \ell^{-n} \hat G\,.
\ee
(The detail derivation is presented in appendix \ref{app:kk}.)

We now consider the reduction of the Kaluza-Klein AdS theory on a $q$-dimensional torus ${\cal T}^q$, keeping only the breathing mode $Y$.  The metric ansatz is
\be
d\tilde s_{\widetilde D}^2 = d\tilde s_{p+2}^2 + Y^2 ds_{{\cal T}^q}^2\,.
\ee
The matter fields in $\tilde D=p+2 + q$ dimensions all descend down directly, giving the Lagrangian
\bea
\widetilde {\cal L}_{p+2} &=& {\cal V}_q\, \sqrt{|g|}\,  Y^q \Bigg(R + q(q-1) Y^{-2} (\partial Y)^2 - \tilde n(\tilde n-1) \Phi^{-2} (\partial \Phi)^2 \nn\\
&&-\fft{1}{\ell^2 \Phi^2}\Big((\tilde n-1)\Phi^{-2\tilde n} -(\tilde n^2-1)\Big)- \fft{1}{4} \Phi^{-2(\tilde n-1)} F^2\Bigg),\label{lowd2}
\eea
where ${\cal V}_q$ is the volume of the torus, implying that the Newton's constant is
\be
\tilde G_{p+2}= {\cal V}^{-1}_q \tilde G \,.
\ee
It is clear that the two Lagrangians (\ref{lowd1}) and (\ref{lowd2}) in $(p+2)$ dimensions have the same structure; in fact, they can map one to the other with the following dictionary
\be
U= \Phi\,,\qquad {\cal A}= A\,,\qquad X= Y^{-1}\,,\qquad
\ell^{-n} \hat G= {\cal V}_q^{-1} \widetilde G\,,
\ee
together with the mapping of the parameters of the dimensions
\be
n\leftrightarrow \tilde n\,,\label{dimensionmap}
\ee
which implies the two parallel and equivalent maps:
\be
q\leftrightarrow -(\hat D-2)\,,\qquad \Leftrightarrow\qquad n\leftrightarrow -(\tilde D-2)\,.\label{dimrelation}
\ee

    It is important to emphasize that the mapping of the dimension parameters (\ref{dimensionmap}) is denoted with
``$\leftrightarrow$'' rather than ``$=$'', since the mapping relates one positive and one negative number and they cannot be equal. Thus the correspondence is not achieved by two different higher-dimensional theories reducing on different internal spaces and yielding the same lower-dimensional theory, as in the typical case of $U$-dualities in string theory. In order for the two theories meeting in dimensions $p+2$, the dimension of one of the internal spaces would have to be negative.  Nevertheless, the mapping provide a way for relating a class Ricci-flat solutions in $\hat D$ dimensions to those of Kaluza-Klein AdS theories in $\widetilde D$ dimensions.

\section{The mapping of solutions}
\label{sec:mapofsolutions}

In this section, we consider two concrete examples of relating classes of Ricci-flat solutions with $U(1)$ bundle of $\mathbb{CP}$ isometries to those in the Kaluza-Klein AdS theory.

\subsection{Ricci-flat Kerr/Charged AdS black hole}

Higher-dimensional generalization of Ricci-flat Kerr metrics \cite{Kerr:1963ud} was obtained in \cite{Myers:1986un}. The general Kerr-AdS black holes have also been constructed \cite{Hawking:1998kw,Gibbons:2004js,Gibbons:2004uw}. We consider Ricci-flat Kerr metrics in odd $n+2$ dimensions with all equal angular momenta, and the metric is cohomogeneity one, given by \cite{Feng:2016dbw}
\be
ds_{n+2}^2 = \fft{dr^2}{f} - \fft{f}{H} dt^2 + r^2 H (\sigma + \omega)^2 + r^2 d\Sigma_{m}^2\,,\label{genkerr}
\ee
where $n=2m+1$ and
\be
f=H- \fft{\mu}{r^{n-1}}\,,\qquad H=1 + \fft{\nu}{r^{n+1}}\,,\qquad
\omega = \fft{\sqrt{\mu\nu}}{r^{n+1} +\nu}dt\,.
\ee
The mass and angular momentum are given by
\be
M= \fft{n\Omega_{n}}{16\pi}\mu\,,\qquad J=\fft{\Omega_{n}}{8\pi} \sqrt{\mu\nu}\,,
\ee
where $\Omega_n=\frac{2\pi^{(n+1)/2}}{\Gamma[\fft12(n+1)]}$ is the volume of the round unit $n$-sphere. The advantage of this parameterization is that it reduces to some new interesting solutions in various limits.  For example, when both $\mu$ and $\nu$ are negative, corresponding to negative mass, the solution becomes a geodesically complete time machine \cite{Feng:2016dbw}, instead of developing a naked singularity, as one might naively expect. When $\nu=0$, the solution reduces to the Schwarzschild black hole in general $n+2$ dimensions. When $\mu=0$, both mass and angular momentum vanish, but the solution remains nontrivial, and we shall discuss this case in the next subsection. The Ricci-flat metric in $\hat D=n+p+2$ dimensions can thus be written as
\be
d\hat s_{\hat D}^2 = \fft{dr^2}{f} - \fft{f}{H} dt^2 + dx^i dx^i +  r^2 H (\sigma + \omega)^2 + r^2 d\Sigma_{m}^2\,,
\ee
where $i=1,2,\ldots, p$.  Following the reduction ansatz \eqref{reduction1}, we obtain the solution in $p+2$ dimensions
\bea
ds_{p+2}^2 &=& \fft{1}{X^2}\Big(\fft{dr^2}{f} - \fft{f}{H} dt^2 + dx^i dx^i\Big)\,,\nn\\
X^2 &=& \fft{r^2}{\ell^2} H^{\fft1{n}}\,,\qquad U=H^{-\fft1{2n}}\,,\qquad {\cal A}=\fft{\sqrt{\mu\nu}}{r^{n+1} +\nu} \ell dt\,.
\eea
Applying the mapping, we obtain
\bea
d\tilde s_{p+2}^2 &=& Y^2 \Big(\fft{dr^2}{\tilde f} - \fft{\tilde f}{\widetilde H} dt^2 + dx^i dx^i\Big)\,,\qquad
Y^2 = \fft{\ell^2}{r^2} \widetilde H^{-\fft1{\tilde n}}\,,\qquad \Phi=\widetilde H^{-\fft1{2\tilde n}}\,,\nn\\
A &=& \fft{\sqrt{\mu\nu}}{r^{\tilde n+1} +\nu}dt\,,\qquad \tilde f=\widetilde H- \fft{\mu}{r^{\tilde n-1}}\,,\qquad \widetilde H=1 + \fft{\nu}{r^{\tilde n+1}}\,.
\eea
Lifting the solution back to $\tilde D$ dimensions, we have the charged AdS black hole in general $\widetilde D$ dimensions.  Making a coordinate transformation $r=\ell^2/\tilde r$, we have
\bea
d\tilde s_{\widetilde D}^2 &=& -\widetilde H^{-\fft{\widetilde D-3}{\widetilde D-2}} \bar f  dt^2 +
\widetilde H^{\fft{1}{\widetilde D-2}} \Big( \fft{d\tilde r^2}{\bar f} + \fft{\tilde r^2}{\ell^2} (dx^i dx^i + ds_{{\cal T}^q}^2)\Big)\,,\nn\\
A &=& \fft{\sqrt{\tilde\mu\tilde \nu}}{\tilde r^{\widetilde D-3} + \tilde \nu} dt\,,\qquad \Phi=\widetilde H^{\fft{1}{2(\widetilde D-2)}}\,,\qquad \widetilde H = 1 + \fft{\tilde \nu}{\tilde r^{\widetilde D-3}}\,,\nn\\
\bar f &=& \fft{\tilde r^2}{\ell^2} \tilde f = \fft{\tilde r^2}{\ell^2} \widetilde H - \fft{\tilde \mu}{\tilde r^{\widetilde D-3}}\,,\qquad \tilde \mu=\mu \ell^{2(\widetilde D-3)}\,,\qquad \tilde \nu = \nu \ell^{2(\widetilde D-2)}\,.
\eea
It can be easily checked that these solve the equations of motion of the theory (\ref{kkemdads}).  In fact, the solution is already in the natural ``$p$-brane coordinates'', and the function $\widetilde H$ is the ``harmonic'' function in the language of \cite{Lu:2013eoa}. The low-lying examples of these solutions can be lifted to describe the decoupling limit of the singly rotating D3-brane or M-branes \cite{Cvetic:1999xp}.

It should be pointed out that the coordinate transformation $r=\ell^2/\tilde r$ is not the necessary part of the mapping, but it puts the solution into the form that was already known. The mass and the electric charge are then given by \cite{Lu:2013eoa}
\be
\widetilde M=\fft{(\widetilde D-2) \Omega_{\widetilde D-2}}{16\pi} \tilde\mu\,,\qquad
\widetilde Q=\fft{(\widetilde D-3) \Omega_{\widetilde D-2}}{16\pi} \sqrt{\tilde \mu\tilde \nu}\,.
\ee
We thus see a mapping of mass and angular momentum in Ricci-flat spacetime to mass and charge in asymptotic AdS geometry
\be
(M,J)\quad \leftrightarrow \quad (\widetilde M, \widetilde Q)\,.\label{mjmq}
\ee
Note that when $\tilde \mu$ and $\tilde \nu$ are both negative, the mass $\widetilde M$ is negative and the solution has naked singularity. The corresponding negative $M$ Ricci-flat Kerr metric however describes a smooth time machine.
Both mass and charge vanish when $\tilde \mu=0$, leading to an AdS domain wall, which we shall discuss next.

\subsection{Gravitational instantons/domain walls}

In the previous section, we establish the correspondence of Ricci-flat metrics to solutions in the Kaluza-Klein AdS theories in general dimensions. If we turn of the Maxwell field, the Einstein-scalar theory would provide an alternative but equivalent ways of constructing Ricci-flat metrics and the metric (\ref{genkerr}) becomes a direct product of time and some Ricci-flat gravitational instanton. In this section, we shall examine the details of the correspondence between these instantons to the domain walls in the Einstein scalar theory.

\subsubsection{Bolt instanton/domain wall}

As was discussed in the previous section, when $\mu=0$, the mass and angular momentum of the metric (\ref{genkerr}) both vanish, but the solution remains nontrivial.  It is a direct product of time and an $(n+1)$-dimensional gravitational instanton, with the metric
\be
ds_{n+1}^2 = \fft{dr^2}{H} + r^2 H \sigma^2 + r^2 d\Sigma_m^2\,,\qquad H=1 + \fft{\nu}{r^{n+1}}\,.\label{geneh}
\ee
These metrics are the Ricci-flat limit of inhomogeneous metrics on complex line bundle \cite{Page:1985bq}, and
can be viewed as one of the higher-dimensional generalizations of the Eguchi-Hanson instanton \cite{Eguchi:1978xp}, or the $\mathbb R^2$ resolved conifold \cite{PandoZayas:2000ctr}.  A different generalization was called the Stenzel metrics \cite{Cvetic:2000db}, which include the deformed conifold in six dimensions \cite{Candelas:1989js}. The metrics can rise from first-order equations using the superpotential formalism that is indicative of reduced holonomy, as discussed in appendix \ref{app:grinst}.

The regularity requires that the parameter $\nu$ is negative so that the cone of $\mathbb{CP}^m$ is blown up to $\mathbb{CP}^m$ at $r_0=(-\nu)^{1/(n+1)}$. The regularity at $r_0$ requires that the period of the $U(1)$ fibre is
\be
\Delta \psi = \fft{4\pi}{n+1}\,,
\ee
which is half required for the level surfaces to be $n$-sphere.  Thus the level surface is $S^n/\mathbb{Z}_2$ and the asymptotic space is $\mathbb{R}^{n+1}/\mathbb{Z}_2$, and the topology of the manifold is $\mathbb{R}^2 \times \mathbb{CP}^m$. Such topology is typically referred to as the ``bolt'' in literature, as opposed to the ``NUT'', which we shall discuss next.

As was discussed in the previous section, this solution can be mapped to the domain wall, which is the massless limit of the charged AdS black hole.  Specifically, the solution is
\bea
ds_{\widetilde D}^2 &=& \fft{\ell^2 d\tilde r^2}{\tilde r^2 \widetilde H^{\fft{\widetilde D-3}{\widetilde D-2}}}+
\fft{\tilde r^2}{\ell^2} \widetilde H^{\fft{1}{\widetilde D-2}} dx^\mu dx^\nu \eta_{\mu\nu}\,,\nn\\
\Phi &=& \widetilde H^{\fft{1}{2(\widetilde D-2)}}\,,\qquad \widetilde H = 1 + \fft{\tilde \nu}{\tilde r^{\widetilde D-3}}\,,
\eea
As we discuss in the appendices \ref{app:dw} and \ref{app:grinst}, both the Ricci-flat bolt instanton and the AdS domain wall can be obtained from the first-order equations {\it via} the superpotential approach.  This suggests that if the theories can be embedded in supergravities, both solutions are supersymmetric, and the correspondence may be established for the fermions as well. Indeed in lower $\widetilde D=7, 6,5,4$ dimensions, the domain wall can be lifted to describe distributed supersymmetric  M-branes or D3-branes, e.g.~\cite{Kraus:1998hv,Russo:1998by,Cvetic:1999xx}. Our correspondence suggests a further connection between the distributed branes and the Ricci-flat gravitational instantons.

\subsubsection{Taub-NUT/domain wall}

Gravitational instantons involving the $U(1)$ bundle over $\mathbb{CP}^m$ also admit the higher-dimensional generalizations of the Taub-NUT solution. The solution in general $D=n+1$ dimensions is given by
\be
ds^2_{n+1}=\fft{dr^2}{f} + f N^2 \sigma^2 + (r^2 - N^2) d\Sigma_m^2\,, \qquad n=2m+1\,,
\ee
where $N$ is the NUT charge. The Einstein metrics of this type was studied in \cite{Page:1985bq,Awad:2000gg}. The Ricci-flat condition implies that the function $f$ satisfies the differential equation
\be
r (r^2-N^2) f' + ((n-2) r^2 + N^2) f - (n+1)(r^2-N^2)=0\,.
\ee
The general solution is given by
\be
f=\Big(1 - \fft{N^2}{r^2}\Big)^{-\ft12(n-1)}\Big(-\fft{c}{r^{n-2}} + \fft{n+1}{n-2}\,
{}_2F_1[\ft12(1-n), \ft12(2-n); \ft12(4-n); \ft{N^2}{r^2}]\Big)\,.
\ee
For the solution to be absent from curvature singularity at $r=N$, the integration constant $c$ is chosen to be
\be
c=
\frac{n \left(\frac{1}{N}\right)^{2-n} \Gamma \left(-\frac{n}{2}\right) \Gamma \left(\frac{n+3}{2}\right)}{\sqrt{\pi }}\,.
\ee
Note that when the dimensional parameter $n$ to be odd integers, the hypergeometric function becomes a rational polynomial.  Here are some low-lying examples:
\bea
n+1=4:&& f= \fft{4(r-N)}{r+N}\,,\nn\\
n+1=6:&& f= \fft{2(r-N)(r+3N)}{(r+N)^2}\,,\nn\\
n+1=8:&& f= \fft{8(r-N) (r^2 + 4Nr + 5N^2)}{5(r+N)^3}\,,\nn\\
n+1=10:&& f= \fft{2(r-N) (5r^3 + 25 N r^2 + 47 N^2 r + 35 N^3)}{7(r+N)^4}\,.\label{nutf}
\eea
({\it A priori}, the mathematical equation is valid for even $n$ as well, in which case, a careful limit has to be taken for the hypergeometric functions.) The metric has a coordinate singularity at $r=N$, which is actually a small patch of $\mathbb{R}^{n+1}$.  To see this explicitly, we note that $f'(N)=2/N$ and thus we let $r-N=\ft14 \rho^2$ and we find that in the limit of $\rho\rightarrow 0$, we have
\be
ds_{n+1}^2 \sim \fft{N}{2} \Big(d\rho^2 + \rho^2 (\sigma^2 + d\Sigma_m^2)\Big).
\ee
Thus the topology of the Taub-NUT is $\mathbb{R}^{n+1}$, very different from the bolt configuration studied earlier.
In the large $r$ limit, $f\rightarrow (n+1)/(n-2)$, the radius of the fibre $\sigma$ is finite whilst the rest ${\cal M}_n$ of the space becomes asymptotically Ricci flat.  However, the ${\cal M}_n$ is not asymptotically flat since it is a foliation of $\mathbb{CP}^m$ not a sphere.  Using the map established earlier, we obtain the new domain wall solution
\bea
d\tilde s_{\widetilde D}^2 &=& \fft{\ell^2}{r^2 - N^2} \Big(\fft{N^2 \tilde f}{r^2 -N^2} \Big)^{\fft{1}{\widetilde D-2}}
\Big(\fft{dr^2}{\widetilde f} + \eta_{\mu\nu} dx^\mu dx^\nu\Big)\nn\\
\Phi &=& \Big(\fft{N^2 \tilde f}{r^2 - N^2}\Big)^{\fft1{2(\widetilde D-2)}}\,,
\eea
where $\tilde f$ satisfy the first-order differential equation
\be
r (r^2-N^2) \tilde f' + (-\widetilde D r^2 + N^2) \tilde f + (\widetilde D-3)(r^2-N^2)=0\,.
\ee
The general solution is $\tilde f = \tilde f_0 + \tilde f_1$, where $\tilde f_0$ is the solution associated with the homogeneous part of the equation and $\tilde f_1$ is a special solution.  We have
\bea
\tilde f_0 &=& -c r^{\widetilde D} \Big(1 - \fft{N^2}{r^2}\Big)^{\fft12(\widetilde D-1)}\,,\nn\\
\tilde f_1 &=& \Big(1 - \fft{N^2}{r^2}\Big)^{\fft12(\widetilde D-1)}\Bigg(\frac{2}{\sqrt{\pi }} \left(\ft{1}{N}\right)^{\widetilde D} \Gamma \left(\ft{5}{2}-\ft{\widetilde D}{2}\right) \Gamma \left(\ft{\widetilde D}{2}\right)r^{\widetilde D}\nn\\
 &&\qquad +  \ft{\widetilde D-3}{\widetilde D} {}_2F_1[
\ft12(\widetilde D-1), \ft12\widetilde D; \ft12 (\widetilde D+2); \ft{N^2}{r^2}]\Bigg)\,,\label{nonsusydwf}
\eea
where $c$ is an integration constant. The special solution $\tilde f_1$ is so arranged such that for even $\widetilde D$, $\tilde f_1$ is polynomial of $r$. Note that we use the original Taub-NUT $r$ coordinate to present the domain wall solution and the asymptotic AdS is located at $r=N$, for which $\tilde f(N)=0$ and $\tilde f'(N)=2/N$. It follows that $\Phi(N)=1$ and $\Phi'(N)=0$.  Some low-lying examples are
\bea
\tilde D=4:&& \tilde f_1=N^{-4}\,(r^2 - N^2)(2r^2-N^2)\,,\nn\\
\tilde D=5:&& \tilde f_1=N^{-5}\, (r^2-N^2) \Big(2 N^3+3 N r^2-3 r \left(r^2-N^2\right) \tanh ^{-1}\left(\frac{r}{N}\right)\Big)\,,\nn\\
\tilde D=6:&& \tilde f_1=N^{-6}\, (r^2-N^2) (-3 N^4+12 N^2 r^2-8 r^4)\,,\nn\\
\tilde D=7:&& \tilde f_1 =\ft12 N^{-7}\, (r^2-N^2)
\Big(-8 N^5+25 N^3 r^2-15 N r^4\nn\\
&&\qquad\qquad+ 15 r \left(r^2-N^2\right)^2 \tanh ^{-1}\left(\frac{r}{N}\right)\Big)\,,\nn\\
\tilde D=8:&& \tilde f_1=N^{-8}\, (r^2-N^2)(-5 N^6 + 30 N^4 r^2 - 40 N^2 r^4 + 16 r^6)\,.
\eea
Note that $\tilde f_1$ in even $\widetilde D$ dimensions maps to the $f$ (\ref{nutf}) of the Taub-NUT in odd dimensions.  Both of the Taub-NUT and domain walls are not solutions of the first order equations {\it via} the superpotential formalism discussed in appendices \ref{app:dw} and \ref{app:grinst}.

A puzzle emerges at this stage in $\widetilde D=4$ dimensions.  This domain wall is related to the four-dimensional Taub-NUT which is supersymmetric, arising from the superpotential formalism when the coefficient $c=0$, as we discussed in appendix \ref{app:grinst}.  On the other hand, the domain wall itself does not satisfy the first order equations (\ref{dwfo}) for the superpotential (\ref{superpot}), namely
\be
W=\fft1{\sqrt2\ell}\Big(3e^{\fft{1}{2\sqrt3}\phi} + e^{-\fft{\sqrt3}2\phi}\Big)\,.
\label{d4sup1}
\ee
This discrepancy can only be resolved that the scalar potential must admit a new superpotential.  Indeed this is the case, and after some careful searching, we arrive at a new unsual looking superpotential
\be
W=\fft{2\sqrt2}{\ell} \cosh^{\fft32} (\ft1{\sqrt3} \phi)\,.\label{d4sup2}
\ee
The domain walls associated with the superpotentials (\ref{d4sup1}) and (\ref{d4sup2}) are thus inspired by the supersymmetric Eguchi-Hanson and Taub-NUT geometry.

    In fact, following from the first order equations (\ref{dwfo}), we can derive the superpotential $W$ using the
known domain wall solution. However, in practice, it involves finding the inversion of the general hypergeometric function in (\ref{nonsusydwf}), which does not have close form expression for general $\widetilde D$.  Another lower-lying example can be written explicitly is $\widetilde D=6$, given by
\be
W=\frac{\left(\sqrt{3-2 e^{2 \sqrt{\frac{2}{5}} \phi }}-2\right) \left(\sqrt{3-2 e^{2 \sqrt{\frac{2}{5}} \phi }}+3\right)^{3/2}}{\sqrt{2}\ell e^{\frac{1}{2} \sqrt{\frac{5}{2}} \phi } }\,.
\ee
It should be pointed that the existence of superpotential does not necessarily imply the supersymmetry.  In fact, it was shown that even the Schwarzschild-AdS black hole can be constructed {\it via} a superpotential formalism \cite{Lu:2003iv}.

Using the reduction ansatz specified in \cite{Cvetic:1999xp}, the domain wall in $\widetilde D=5$ can be lifted to describe certain non-supersymmetric D3-brane deformation, with the ingredients, in notation of \cite{Cvetic:1999xp},
\be
X_1=\Phi^{-4},\qquad X_2=X_3=\Phi^2\,,\qquad \tilde{\Delta}=\Phi^{-4}\mu_1^2+\Phi^2(1-\mu_1^2)\,,\qquad
g^2=\frac{1}{\ell^2}\,.
\ee
The deformation of M2-brane can be obtained from the uplift of $\widetilde D=4$ domain wall, with
\be
X_1=\Phi^{-3},\qquad X_2=X_3=X_4=\Phi\,,\qquad
\tilde{\Delta}=\Phi^{-3}\mu_1^2+\Phi(1-\mu_1^2)\,,\qquad
g^2=\frac{1}{4\ell^2}\,.
\ee
Finally the deformation of M5-brane can be obtained from the uplift of $\widetilde D=7$ domain wall, with
\be
X_1=\Phi^{-6},\qquad X_0=X_2=\Phi^4\,,\qquad
\tilde{\Delta}=\Phi^{-6}\mu_1^2+\Phi^4(1-\mu_1^2)\,,\qquad
g^2=\frac{4}{\ell^2}\,.
\ee
These deformations  are not described by the supersymmetric distributed branes and are not supersymmetric.

\section{Conclusion}

In this paper, we studied the AdS/Ricci-flat correspondence proposed in \cite{Caldarelli:2013aaa} and
established an explicit map between Einstein gravity on squashed spheres that are the $U(1)$ bundles over $\mathbb{CP}^m$ and Kaluza-Klein AdS gravity on tori. In dimensions $\widetilde D=7,6,5,4$, the Kaluza-Klein AdS theory can be embedded in gauged supergravities that can be obtained from consistent sphere reductions of M-theory and type IIB supergravity. This not only allows to relate the previously known Ricci-flat solutions to brane solutions, but also yield new deformations of M-brane and D3-branes.

Specifically, we found that Ricci-flat Kerr metrics with all equal angular momenta in odd dimensions can map to charged Kaluza-Klein AdS black holes, with the mapping of the conserved quantities (\ref{mjmq}).  In suitable dimensions, the charged black hole can be lifted to become singly rotating M-branes and D3-branes.  What is intriguing is that the correspondence occurs between the multiply rotating Ricci-flat metrics and singly rotating branes.  We also found maps between gravitational instantons to AdS domain walls.  In particular, the supersymmetric bolt instantons relates to the supersymmetric domain walls that have origins of distributed M-branes or D3-branes.  The non-supersymmetric Taub-NUT solutions yield new domain walls that can lifted to become new deformations of the AdS$\times$sphere vacua of M-theory and type IIA supergravity.

It should be pointed that the correspondence involves negative dimensions, as in (\ref{dimrelation}).  This implies that the correspondence can be only established for solutions in general dimensions where the dimension is a parameter. If we have a solution in a specific dimension only, there is no obvious way to set up the transformation rules from one theory to the other. Nevertheless, the correspondence is quite powerful, allows us even to obtain new solutions in M-theory and type IIB supergravity.  In fact, it encouraged us to find a new superpotential for the scalar potential in the Kaluza-Klein AdS supergravity in four dimensions and it does exist! These give a tantalizing suggestion that the concept of negative spacetime dimensions may not be simply dismissed. It is also of great interest to extend our discussion to include more general sphere distortions. The correspondence may also shed light on the holography between Minkowski spacetimes and quantum field theory.

\section*{Acknowledgement}

We are grateful to Chris Pope for useful discussions and to Rui Wen who gave a clear invited talk in CJQS on the AdS/Ricci-flat correspondence last winter.  The work is supported in part by NSFC (National Natural Science Foundation of China) Grants No.~11875200 and No.~11935009.

\appendix

\section{Kaluza-Klein reduction and curvature tensor}
\label{app:kk}

In this appendix, we perform Kaluza-Klein reduction of Einstein gravity on the squashed sphere that is the $U(1)$ bundle over $\mathbb{CP}^m$. The metric ansatz is
\bea
d\hat{s}_{\hat D}^2 &=&d\hat{s}_{p+2}^2+e^{2\beta_1\phi_1}(\sigma+\ell^{-1}\mathcal{A})^2+e^{2\beta_2\phi_2}d\Sigma_m^2
\eea
The natural choice of vielbein is $(e^a,e^{\sigma},e^i)$,\ with
\bea
e^{\sigma}=e^{\beta_1\phi_1}(\sigma+\ell^{-1}\mathcal{A}), \ \ e^{i}=e^{\beta_2\phi_2}e_\Sigma^{i}\,,
\eea
where $a=0,1,\ldots, p+1$ and $i=\bar 1, \bar 2, \ldots, \bar {2m}$. We follow the identical convention of the $\mathbb{CP}$ metrics presented in the appendix of \cite{Cvetic:2018ipt}, we thus have
\bea
d\sigma=2J, \ \ J=\ft{1}{2}J_{ij}e_\Sigma^{i}\wedge e_\Sigma^{j},\ \ \ \mathcal{F}=d\mathcal{A} = \ft{1}{2}\mathcal{F}_{ab}e^a\wedge e^b
\eea
The exterior derivatives of the vielbein are
\bea
de^a =&&-\overline{\omega}^a{}_{b}\wedge e^b,\cr
de^{\sigma} =&&\beta_1 \partial_b\phi_1 e^b\wedge e^{\sigma}+e^{\beta_1\phi_1-2\beta_2\phi_2}J_{ij}e^{i}\wedge e^{j}+\ell^{-1}e^{\beta_1\phi_1}\mathcal{F},\cr
de^{i} =&&-\omega_\Sigma^{i}{}_{j}\wedge e^{j}+\beta_2\partial_b\phi_2e^b \wedge e^{i},
\eea
where $\overline{\omega}^a{}_{b}$ and $\omega_\Sigma^{i}{}_{j}$ are the spin connections of the lower dimension spacetime $d\hat{s}_{p+2}^2$ and $\mathbb{CP}^m$ respectively. From these, we obtain the non-vanishing components of the spin connection
\bea
\omega^{\sigma}{}_a &=&\beta_1\partial_a\phi_1e^{\sigma}+\ft{1}{2}\ell^{-1}e^{\beta_1\phi_1}\mathcal{F}_{ab}e^b,\qquad
\omega^{\sigma}{}_{i} =e^{\beta_1\phi_1-2\beta_2\phi_2}J_{ij}e^{j},\cr
\omega^{i}{}_b &=&\beta_2\partial_b\phi_2e^{i},\qquad
\omega^{i}{}_{j} =\omega_{\Sigma}^{i}{}_{j}-e^{\beta_1\phi_1-2\beta_2\phi_2}J^{i}{}_{j}e^{\sigma},\qquad
\omega^a{}_b =\overline{\omega}^{a}{}_{b}-\ft{1}{2}\ell^{-1}e^{\beta_1\phi_1}\mathcal{F}^a{}_{b}e^{\sigma}\,.
\eea
The components of the 2-form curvature tensor $\Theta=d\omega + \omega\wedge \omega$ are thus given by
\bea
\Theta^{\sigma}{}_{a} =&&\left[\beta_1\nabla_b\nabla_a\phi_1+\beta_1^2(\nabla_a\phi_1)(\nabla_b\phi_1)\right]e^b
\wedge e^{\sigma}-\ft{1}{4}\ell^{-2}e^{2\beta_1\phi_1}\mathcal{F}_{bc}\mathcal{F}^{b}{}_{a}e^c\wedge e^{\sigma}\cr
&&+(\beta_1\nabla_a\phi_1-\beta_2\nabla_a\phi_2)e^{\beta_1\phi_1-2\beta_2\phi_2}J_{ij}e^{i}\wedge e^{j}\cr
&&+\ell^{-1}\beta_1(\nabla_a\phi_1)e^{\beta_1\phi_1}\mathcal{F}+\ft{1}{2}\ell^{-1}e^{\beta_1\phi_1}
\left[\beta_1(\nabla_b\phi_1)\mathcal{F}_{ac}+\nabla_b \mathcal{F}_{ac}  \right]e^b\wedge e^c,\cr
\Theta^{\sigma}{}_{i}=&&(\beta_1\nabla_c\phi_1-\beta_2\nabla_c\phi_2)e^{\beta_1\phi_1-2\beta_2\phi_2}J_{ij}e^{c}
\wedge e^{j}-\ft{1}{2}\ell^{-1}\beta_2(\nabla^b\phi_2)e^{\beta_1\phi_1}\mathcal{F}_{bc}e^c\wedge e^{k}\delta_{ki}\cr
&&+\beta_1\beta_2(\nabla_b\phi_1)(\nabla^b\phi_2)\delta_{ki}e^{k}\wedge e^{\sigma}-e^{2(\beta_1\phi_1-2\beta_2\phi_2)}J^{j}{}_{i}J_{jk}e^{k}\wedge e^{\sigma},\cr
\Theta^{i}{}_{b} =&&\left[\beta_2\nabla_c\nabla_b\phi_2 +\beta_2^2(\nabla_b\phi_2)(\nabla_c\phi_2)\right]e^c
\wedge e^{i}-\ft{1}{2}\ell^{-1}e^{2(\beta_1\phi_1-\beta_2\phi_2)}\mathcal{F}_{bc}J^{i}{}_{k}e^{k}\wedge e^c\cr
&&-\ft{1}{2}\ell^{-1}\beta_2(\nabla_c\phi_2)e^{\beta_1\phi_1}\mathcal{F}^{c}{}_{b}e^{i}\wedge e^{\sigma}-(\beta_{1}\nabla_b\phi_1-\beta_{2}\nabla_b\phi_2)e^{\beta_1\phi_1-2\beta_2\phi_2}J^{i}{}_{k}e^{k}\wedge e^{\sigma},\cr
\Theta^{i}{}_{j} =&&d\omega_{\Sigma}^{i}{}_{j}+\omega_{\Sigma}^{i}{}_{k}\wedge \omega_{\Sigma}^{k}{}_{j}-e^{2(\beta_1\phi_1-2\beta_2\phi_2)}J^{i}{}_{j}J_{kl}e^{k}\wedge e^{l}\cr
&&-\beta_2^2(\nabla\phi_2)^2\delta_{jk}e^{i}\wedge e^{k}-e^{2(\beta_1\phi_1-2\beta_2\phi_2)}J^{i}{}_{k}J_{jl}e^{k}\wedge e^{l}\cr
&&-2(\beta_{1}\nabla_c\phi_1-\beta_{2}\nabla_c\phi_2)e^{\beta_1\phi_1-2\beta_2\phi_2}J^{i}{}_{j}e^{c}\wedge e^{\sigma}-\ell^{-1}e^{2(\beta_1\phi_1-\beta_2\phi_2)}J^{i}{}_{j}\mathcal{F},\cr
\Theta^a{}_{b} =&&d\overline{\omega}^{a}{}_{b}+\overline{\omega}^{a}{}_{c}\wedge \overline{\omega}^{c}{}_{b}-\ft{1}{2}\ell^{-2}e^{2\beta_1\phi_1}\mathcal{F}^{a}{}_{b}
\mathcal{F}-\ft{1}{4}\ell^{-2}e^{2\beta_1\phi_1}\mathcal{F}^{a}{}_{c}\mathcal{F}_{bd}e^c\wedge e^d\cr
&&+\ft{1}{2}\ell^{-1}\beta_1e^{\beta_1\phi_1}\left[(\nabla^a\phi_1)\mathcal{F}_{bc}-
(\nabla_b\phi_1)\mathcal{F}^{a}{}_{c}-2(\nabla_c\phi_1)\mathcal{F}^{a}{}_{b}  \right]e^c\wedge e^{\sigma}\cr
&&-\ft{1}{2}\ell^{-1}e^{\beta_1\phi_1}(\nabla_c\mathcal{F}^{a}{}_{b})e^c\wedge e^{\sigma} -\ft{1}{2}\ell^{-1}e^{2(\beta_1\phi_1-\beta_2\phi_2)}\mathcal{F}^{a}{}_{b}J_{kl}e^{k}\wedge e^{l}.
\eea
We can read off Riemann tensor components using $\Theta^M{}_N=\fft12 R^M{}_{NPQ} e^{P}\wedge e^{Q}$. We split the Riemann tensor components in two parts: those that contribute to the Ricci scalar and those that do not. The latter ones are
\bea
R^{\sigma}{}_{aij}=&&2(\beta_{1}\nabla_a\phi_1-\beta_{2}\nabla_a\phi_2)e^{\beta_1\phi_1-2\beta_2\phi_2}J_{ij},\cr
R^{\sigma}{}_{acd}=&&\ft{1}{2}\ell^{-1}\beta_1e^{\beta_1\phi_1}\left[2(\nabla_a\phi_1)\mathcal{F}_{cd}-
(\nabla_c\phi_1)\mathcal{F}_{da}-(\nabla_d\phi_1)\mathcal{F}_{ac}   \right]\cr
&&+\ft{1}{2}\ell^{-1}e^{\beta_1\phi_1}\nabla_{a}\mathcal{F}_{cd},\cr
R^{\sigma}{}_{icj}=&&(\beta_{1}\nabla_c\phi_1-\beta_{2}\nabla_c\phi_2)e^{\beta_1\phi_1-2\beta_2\phi_2}J_{ij}
-\ft{1}{2}\ell^{-1}\beta_2e^{\beta_1\phi_1}(\nabla^b\phi_2)\mathcal{F}_{bc}\delta_{ij},\cr
R^{i}{}_{jcd}=&&-\ell^{-1}e^{2(\beta_1\phi_1-\beta_2\phi_2)}J^{i}{}_{j}\mathcal{F}_{cd}\,.
\eea
The former ones are
\bea
R^{\sigma}{}_{a\sigma b} =&&-\beta_1\nabla_b\nabla_a\phi_1-
\beta_1^2(\nabla_a\phi_1)(\nabla_b\phi_1)+\ft{1}{4}\ell^{-2}e^{2\beta_1\phi_1}\mathcal{F}^{c}
{}_{a}\mathcal{F}_{cb},\cr
R^{\sigma}{}_{i\sigma j} =&&-\beta_1\beta_2(\nabla_b\phi_1)(\nabla^b\phi_2)\delta_{ij}+e^{2(\beta_1\phi_1-2\beta_2\phi_2)}J^{k}{}_{i}J_{kj},\cr
R^{i}{}_{ajb} =&&-\beta_2\nabla_b\nabla_a\phi_2\delta^{i}_{j}-
\beta^2_2(\nabla_a\phi_2)(\nabla_b\phi_2)\delta^{i}_{j}-\ft{1}{2}\ell^{-1}e^{2(\beta_1\phi_1-\beta_2\phi_2)}J^{i}{} _{j}\mathcal{F}_{ab},\cr
R^{i}{}_{jkl} =&&e^{-2\beta_2\phi_2}R_{\Sigma}^{i}{}_{jkl}-2e^{2(\beta_1\phi_1-2\beta_2\phi_2)}J^{i}{}_{j}J_{kl}\cr
&& -\beta_2^2(\nabla\phi_2)^2(\delta_{jl}\delta^{i}_{k}-\delta_{jk}\delta^{i}_{l})-
e^{2(\beta_1\phi_1-2\beta_2\phi_2)}(J_{jl}J^{i}{} _{k}-J_{jk}J^{i}{}_{l}),\cr
R^{a}{}_{bcd} =&&R_{p+2}^{a}{}_{bcd}-\ft{1}{2}\ell^{-2}e^{2\beta_1\phi_1}\mathcal{F}^{a}{}_{b}\mathcal{F}_{cd}
-\ft{1}{4}\ell^{-2}e^{2\beta_1\phi_1}(\mathcal{F}^{a}{}_{c}\mathcal{F}_{bd}-\mathcal{F}^{a}{}_{d}\mathcal{F}_{bc})\,.
\eea
Here $R_{\Sigma}^{i}{}_{jkl}$ is the Riemann tensor of the internal space $\mathbb{CP}^m$. When $m=0$, the Kaluza-Klein ansatz reduces to that of $S^1$ reduction and the Riemann tensor was given in \cite{Liu:2012ed}. The Ricci tensors are
\bea
R^{a}{}_{b} =&&R_{p+2}^{a}{}_{b}-\ft{1}{2}\ell^{-2}e^{2\beta_1\phi_1}\mathcal{F}^{ac}\mathcal{F}_{bc}
-\beta_1\nabla_b\nabla^a\phi_1-\beta_1^2(\nabla^a\phi_1)(\nabla_b\phi_1)\cr
&& -d\beta_2\nabla_b\nabla^a\phi_2-d\beta_2^2(\nabla^a\phi_2)(\nabla_b\phi_2),\cr
R^{i}{}_{j} =&&e^{-2\beta_2\phi_2}R_{\Sigma}^{i}{}_{j}-\beta_2\Box\phi_2\delta^{i}_{j}-d\beta^2_2(\nabla\phi_2)^2\delta^{i}_{j}\cr
&& -\beta_1\beta_2(\nabla_b\phi_1)(\nabla^b\phi_2)\delta^{i}_{j}-2e^{2(\beta_1\phi_1-2\beta_2\phi_2)}J^{ik}J_{jk},\cr
R^{\sigma}{}_{\sigma} =&&-\beta_1\Box\phi_1-\beta_1^2(\nabla\phi_1)^2-d\beta_1\beta_2(\nabla_b\phi_1)(\nabla^b\phi_2)\cr
&& +\ft{1}{4}\ell^{-2}e^{2\beta_1\phi_1}\mathcal{F}^2+e^{2(\beta_1\phi_1-2\beta_2\phi_2)}J^{kl}J_{kl},
\eea
where $d=2m$ is the dimension of $\mathbb{CP}^m$, together with
\be
R^{\sigma}{}_{a}= \ft{1}{2}\ell^{-1}e^{-2\beta_1\phi_1}\nabla^{c}(e^{3\beta_1\phi_1}\mathcal{F}_{ac})
+\ft{1}{2}\ell^{-1}d\beta_2e^{\beta_1\phi_1}(\nabla^c\phi_2)\mathcal{F}_{ac}\,,
\ee
that does not contribute to the Ricci scalar. The Ricci scalar is
\bea
R &=&R_{p+2}-\ft{1}{4}\ell^{-2}e^{2\beta_1\phi_1}{\cal F}^2-2\beta_1\Box\phi_1-2\beta_1^2(\nabla\phi_1)^2\cr
&& -2d\beta_1\beta_2(\nabla_b\phi_1)(\nabla^b\phi_2)-2d\beta_2\Box\phi_2-d(d+1)\beta_2^2(\nabla\phi_2)^2\cr
&& +e^{-2\beta_2\phi_2}R_{\Sigma}-e^{2(\beta_1\phi_1-2\beta_2\phi_2)}J^{ij}J_{ij}.
\eea
After integrating by part, the action becomes
\bea
\mathcal{L}_{p+2} =&&e^{(\beta_1\phi_1+d\beta_2\phi_2)}\Big(R_{p+2}-\ft{1}{4}\ell^{-2}e^{2\beta_1\phi_1}{\cal F}^2+d(d-1)\beta_2^2(\nabla\phi_2)^2\cr
&& +2d\beta_1\beta_2(\nabla_b\phi_1)(\nabla^b\phi_2)-e^{2(\beta_1\phi_1-2\beta_2\phi_2)}J^{ij}J_{ij}+e^{-2\beta_2\phi_2}R_{\Sigma}\Big).
\eea
If we set
\bea
\phi_1 &=&\frac{1}{\beta_1}\log\Big(\frac{X\ell}{U^{(n-1)}}\Big),\ \ \phi_2 = \frac{1}{\beta_2}\log(UX\ell),\ \ d=n-1,
\eea
the cross term\ $2d\beta_1\beta_2(\nabla_b\phi_1)(\nabla^b\phi_2)$\ will be remove and the action becomes to
\bea
\mathcal{L}_{p+2} =&&\ell^nX^n\sqrt{|\hat g|}\Big(R_{p+2}-\frac{1}{4}\frac{X^2}{U^{2(n-1)}}{\cal F}^2+n(n-1)X^{-2}(\partial X)^2\cr
&& -n(n-1)U^{-2}(\partial U)^2-\frac{n-1}{\ell^2X^2U^{2(n+1)}}+\frac{n^2-1}{\ell^2X^2U^2}
\Big).
\eea
Notice that in the convention of \cite{Cvetic:2018ipt} we have
\bea
J^{ij}J_{ij} &=&n-1,\ \ \ R_{\Sigma} = n^2-1\,.
\eea
After a further conformation transformation
\be
d\hat s_{p+2}^2 = X^2 ds_{p+2}^2\,,
\ee
we arrive at the Lagrangian (\ref{lowd1}).

\section{Domain wall solution}
\label{app:dw}

In this paper, we consider $\widetilde D$-dimensional domain walls whose world-volume is the Minkowski spacetime and the ansatz is
\be
ds^2_{\widetilde D}=e^{2B(r)} dr^2 + e^{2A(r)} dx^\mu dx^\nu\eta_{\mu\nu}\,,\qquad \phi=\phi(r)\,.
\ee
For the Einstein-scalar theory (\ref{kkemdads}) (with $F=0$), the equations of motion are
\bea
\phi'' + \Big((\widetilde D-1) A' - B'\Big)\phi'-e^{2B} \fft{dV}{d\phi}&=&0\,,\nn\\
(\widetilde D-2) A'' -(\widetilde D-2) A'B' + \ft12 \phi'^2 &=& 0\,,\nn\\
A''+\Big((\widetilde D-1) A' - B'\Big) A' + \fft{e^{2B}}{\widetilde D-2}V &=&0\,.
\eea
When the scalar can be expressed in terms of a superpotential, as in (\ref{superpot}), there exist ``BPS'' solutions that satisfy the first order equations
\be
\phi' = \mp \sqrt2 e^{B} \fft{dW}{d\phi}\,,\qquad
A'=\pm\fft{e^B W}{\sqrt{2(\widetilde D-2)}}\,.\label{dwfo}
\ee
Note that the $\pm$ sign choice is trivial and can be absorbed into the $r$ coordinate.

\section{Superpotential formalism of the gravitational instantons}
\label{app:grinst}

In this appendix, we consider a class of cohomogeneity-one Ricci-flat gravitational instanton whose
level surfaces are described by a $U(1)$ bundle over $\mathbb{CP}^m$.  The metric ansatz in general $D=n+1$ dimensions is
\be
ds^2 = a^2 b^{2m} dr^2 + a^2 \sigma^2 + b^2 d\Sigma_m^2\,,
\ee
where $n=2m+1$ and $(a,b)$ are functions of the radial coordinate $r$.  The system reduces to the Newtonian mechanics and the effective Lagrangian that ensures the Ricci flatness is
\be
L=T - V\,,\qquad T=\fft{2a' b'}{ab} + (D-3) \fft{b'^2}{b^2}\,,\qquad V=a^2 b^{2(D-4)}(a^2 - D b^2)\,.
\ee
Following the technique outlined in \cite{Cvetic:2000db}, we define $a=e^{\alpha_1}$ and $b=e^{\alpha_2}$.  The kinetic term can be
expressed as $T=\ft12 g_{ij} (\alpha^i)' (\alpha^j)'$.  We find that there exists a superpotential
\be
W=\fft{1}{D-2}\Big((D-2) a^2 + D b^2\Big)b^{D-4}\,,
\ee
such that $V=-\ft12 g^{ij} \partial_i W\partial_j W$.  This leads to first order integrals $(\alpha^i)' = g^{ij} \partial_j W$, {\it i.e.}
\be
b' = a^2 b^{D-3}\,,\qquad a'= \ft12 a b^{D-4} \Big(D b^2 - 2(D-2) a^2\Big)\,.
\ee
In some suitable radial coordinate, these two equations can be solved exactly, giving precisely the metric (\ref{geneh}) with $D=n+1$. The $m=1$ case gives the Eguchi-Hanson instanton \cite{Eguchi:1978xp}.  The general solution is a special (BPS) case of inhomogeneous Einstein metrics on complex line bundle studied in \cite{Page:1985bq}. Note that in $D=4$ dimensions, there is an additional superpotential $W=a (a + 4 b)$, which yield the Ricci-flat Taub-NUT solution in four dimensions, corresponding to $n=3$ in (\ref{nutf}).

\end{document}